\documentstyle[preprint,revtex,eqsecnum]{aps}
\begin{document}
\preprint{IP/BBSR/92-60}
\preprint{July, 1992 $\qquad$}
\begin{title}
Phase Transitions in Quantum Field Theory
\end{title}
\author{S.P. Misra}
\begin{instit}
Institute of Physics, Bhubaneswar-751005, India.
\end{instit}
\begin{abstract}
We discuss here phase transitions in quantum field theory in the context of
vacuum realignment through an explicit construction. Vacuum
destabilisation may occur through a scalar attaining a nonzero expectation
value, or through a condensate mechanism as in superconductivity and for
chiral symmetry breaking, or by some other mechanism not taken here. Phase
transition is defined as vacuum realignment with a discontinuity in
some observables which correspond to the classical order parameters.
The method is nonperturbative and variational,
using equal time algebra and an ansatz for the construction of the
physical vacuum.
\end{abstract}
\vfil
\noindent
Invited talk in the Symposium on Statistical Mechanics and Quantum
Field Theory, Calcutta, January, 1992.
\section{\bf {Introduction}}

\bibcite{higgs}{1}
\bibcite{wein}{2}
\bibcite{grsnv}{3}
\bibcite{spm87}{4}
\bibcite{hm88}{5}
\bibcite{hgs}{6}
\bibcite{sn90}{7}
\bibcite{am91}{8}
Symmetries as well as symmetry breaking involving `phase transitions'
are important concepts in physics. Phase transition in quantum field
is usually considered by minimising a {\it classical} potential \cite{higgs}.
Also, for dynamical symmetry breaking, we may include quantum corrections
to the potential, and again minimise it at the classical level
\cite{wein,grsnv}.  In the present talk, we shall consider an alternative
approach through vacuum realignment \cite{spm87,hm88,hgs,sn90,am91}.

\bibcite{glstn}{9}
\bibcite{nambu}{10}
\bibcite{hm92}{11}
The talk is organised as follows. We shall first consider in Section II
the definition of phase transition in quantum field theory \cite{hgs}.
We shall next consider in Section III some simple examples of the same
from quantum mechanics \cite{spm87}. In Section IV we consider
the Gross-Neveu model \cite{grsnv,hm88} as one of the simplest 1+1
dimensional field theoretic models to illustrate the case. In Section V
we consider Goldstone theorem \cite{glstn} in the context of
phase transition with the breaking of a global symmetry. We use this picture
in Section VI to consider chiral symmetry breaking with Nambu
Jona Lasinio model \cite{nambu}
to construct the pion wave function, which constrains, and is constrained
by the vacuum structure \cite{hm92}.
In Section VII we consider gluon condensates to bring out the features
of phase transion in quantum chromodynamics at zero temperature as well
as at finite temperature \cite{am91}. Section VIII consists of a description of
conventional phase transition as a vacuum destabilisation, and, the
corresponding possibility of the production of Higgs particles with
nonperturbative dynamics associated with some high energy cosmic ray
events \cite{hgs}.
In Section IX we discuss the results.

The objective of the present review is to
give a few examples to illustrate the present ideas in relation to
the conventional picture as well as bring
out the extra conceptual and experimental consequences of the
same. The method is nonperturbative, and uses equal time
algebra and a variational ansatz for the construction of the vacuum state
\cite{hgs,am91}.

\section{\bf {Quantum Phase Transition}}
Quantum field theory, or quantum mechanics, start with
the construction of a vacuum state, and then a Fock space of the
observable quantum modes which interact with each other. For this construction,
the fiducial state is the vacuum state, $|vac>$, which we shall some times
call as the {\it base state} over which the excitations are defined.
We start with a different word since the reference state we start with may
not correspond to the physical vacuum. In fact, when we start with an
arbitrary basis, we shall see that the vacuum state
is {\it not} structureless or empty in this basis. It will depend on the
parameters of the theory such as
coupling constants, as well as temperature when we consider quantum
field theory at finite temperature. The base state or `vacuum structure'
will be determined through a minimisation of energy at zero temperature,
or of free energy or thermodynamic potential at finite temperature.
In a given approximation scheme,
we may determine this to the best of our ability.
We shall however presume that it exists, as is a prerequisite for
the construction of Fock space in field theory.

The change in the base state resulting from a change of the parameters
of the theory or of temperature will be called as {\it vacuum
destabilisation}, or, {\it vacuum realignment}, which, as stated
merely means a change of the basis. If this change is
discontinuous, we call it as a {\it quantum phase transition}.
Discontinuity of the above state is defined as discontinuity
in the expectation values of some operators of the theory with respect to the
vacuuum state. These expectation values are in fact the order parameters
for the classical picture of phase transitions and thus a phase
transition depicts a discontinuity of some observables in these parameters.
The order of the phase transition can also be correspondingly defined.

We shall now mathematically formulate the above statements.

Let $|vac>$ be the base state corresponding to the free field vacuum.
When interactions and/or temperature is there, let the base state be
$|vac'>$. Let us substitute \cite{hgs}
\begin{equation}
|vac'>=U|vac>,
\end{equation}
where $U$ is formally a unitary operator since the above should only imply
a redefinition of the basis. As stated earlier, the operator $U$ shall get
determined by minimising energy for zero temperature field theory, and
by minimising free energy or the thermodynamic potential if the
temperature is different from zero. In field
theory we generally wish to have a translationally invariant description.
Hence we minimise over energy density, or free energy density. There is
generally no reason why $U$ should be identity.
Change of $|vac'>$  in a discontinuous manner implies a change
of $U$ in a discontinuous manner. We can determine $U$ and estimate the
appropriate order parameters, and then recognise whether a phase transition
has taken place, as well as the order of the phase transition. This will
be the description for quantum phase transitions, which shall be relevant
with or without symmetry breaking. Here we shall concentrate on
the same particularly in the context of symmetry breaking.

In field theory, with
translational invariance, the operator $U$ mostly has a formal
definition, and does not exist in the strict mathematical sense. We shall
imagine that we can have a regularisation procedure so that the results
formally obtained continue to be true. For example, this can be done through
taking a finite volume, and then in the limit the volume can be made to
go to infinity.

Let us next consider the way we shall be implementing the above. We shall
start with a Lagrangian where quantisation with equal time algebra is
given. We then expand the field operators using the equal time algebra to
define creation and annihilation operators, and therefore a Hilbert space,
where some operators, named as annihilation operators,
annihilate the vacuum. We do not use the free field equations,
and thus in the expansion of the field operators, there are expressions
which are not determined. We next calculate the expectation value of energy
density in terms of these unknown expressions. This shall be
done in a self-consistent manner as illustrated below. The field equations
are here replaced through a variation at the level of expectation values.

\bibcite{feynman}{12}
Vacuum state corresponding to the expansion of the field operators in the
form of free field operators will be known as the perturbative vacuum, and
the basis as the perturbative basis.  If the the arbitrary functions as stated
above are left undetermined, the corresponding basis will be called a general
basis. When we determine the basis through a minimisation of
energy density (or a maximisation of the Lagrangian density), this will
be known as the optimised basis and will correspond to the physical
vacuum. The method is nonperturbative because the basis here will get
determined through essentially a nonperturbative calculation.
We thus throughout concentrate on the operator formulation of quantum
field theory instead of the path integral formulation \cite{feynman}, since
we find that this often has a richer structure of physics output.
We do this fully recognising the limitations of this formulaion.

Besides discussing vacuum realignment, we shall also have occasion to specify
different phases. The phase corresponding to the perturbative basis will
be known as the perturbative phase. The phase as determined through
the minimisation of energy density (or maximising the Lagrangian density)
will be known as the physical phase, and shall correspond to {\it true}
vacuum. Clearly the Hilbert space will be defined over the physical vacuum.
As stated, the change of basis for the physical vacuum with respect to the
parameters of the theory, including temperature, may occur either continuously
or may have a discontinuity. A discontinuos change implies a phase transition.
The critical values of the parameters are those values where this
discontinuity occurs.

We shall now consider different examples of vacuum destabilisation with
a smooth change, as well as of phase transitions where the change is not
smooth.

\section{\bf Quantum Mechanics}
We shall consider here how for the variation of the coupling
constant, we can have a phase transition from a bosonic vacuum to a
fermionic vacuum as a simple example in quantum mechanics \cite{spm87}.

Let us consider the Hamiltonian
\begin{equation}
H=\epsilon c^\dagger c + \omega a ^\dagger a +g(c^\dagger c)(a^\dagger+a)
\end{equation}
with the quantisation condition $[c,c^\dagger]_+=[a,a^\dagger]=1$.
This corresponds
to quantum electrodynamics in (1,0) dimensions with $|vac>$ defined
through $a|vac>=c|vac>=0$. The above Hamiltonian has eigenstates with
no fermions and an arbitrary number of bosons, as well as eigenstates
with a single fermion and an arbitrary number of bosons \cite{spm87}.
We further have through the present construction
\begin{equation}
H|vac>=0
\end{equation}
so that $|vac>$ is the ground state of the bosonic sector.

We next consider the fermionic sector. Clearly, the fermion will be always
dressed with bosons. We may recall that for coherent states, for any $z$,
\begin{equation}
a\; e^{z a^\dagger }|vac>=z\; e^{z a^\dagger }|vac>.
\end{equation}
With $z=-g/\omega$, we then have
\begin{equation}
H\;c^\dagger  e^{-(g/\omega)a^\dagger }|vac>=\left(\epsilon -g^2/\omega\right)
c^\dagger  e^{-(g/\omega)a^\dagger }|vac>.
\end{equation}
This is the eigenvalue equation for the lowest eigenvalue of the fermionic
sector with the energy eigenvalue given as \cite{spm87}
\begin{equation}
E_1=\epsilon-g^2/\omega.
\end{equation}
We then clearly note that for $g^2 > g_c^2 \equiv \epsilon \omega $, there is a
phase transition from the bosonic vacuum with zero energy to a fermionic vacuum
with negative energy. Since the fermion number changes discontinuously, this
is a phase transition. $g_c$ is the critical coupling constant.
The other eigenvalues, which are larger, can also be evaluated \cite{spm87},
and shall be needed if we consider phase transitions at finite temperature.

\bibcite{anderson}{13}
This simple model illustrates the use of coherent states for
discussing exact solutions with a natural vacuum realignment.
We may note that coherent states were used a long time back by Anderson
\cite{anderson} in superconductivity.

\section{\bf Gross-Neveu Model}
We shall consider here as a simple application of the nonperturbative
variational method to Gross Neveu model \cite{grsnv}, which is exactly
solvable in 1/N approximation as a test case \cite{hm88}.

The Lagrangian describing the Gross-Neveu model is given as \cite{grsnv}
\begin{equation}
{\cal L}= i{\bar \psi}\gamma^{\mu}\partial_{\mu}\psi-g{\bar \psi}\psi\sigma
-{1\over 2}\sigma^{2}.
\end{equation}
We expand the interacting field operators at t=0 as
\begin{equation}
\psi(x,0)={1\over \sqrt {2\pi}} \int\left[U(k)c(k)e{^{ikx}} +
V(k){\tilde c}(k)e{^{-ikx}}\right] dk,
\end{equation}
where c and $\tilde c$ are particle annihilation and
anti-particle creation operators satisfying the algebra
\begin{equation}
\left[c(k),c(k^{\prime})^{\dagger}\right]_{+}=
\left[\tilde c(k),\tilde c{(k^{\prime})}^{\dagger}\right]_{+}=
\delta (k-{k^{\prime}}).
\end{equation}
This becomes consistent with
\begin{equation}
\left[\psi_{i}(x,t), \psi_{i}(y,t)^{\dagger}\right]_{+}
=\delta_{i,j}\delta(x-y),
\end{equation}
provided that \cite{hm88}
\begin{equation}
U(k)=\pmatrix{\cos{\phi(k)\over 2}\cr
-i\sin{\phi(k)\over 2}};\qquad V(k)=\pmatrix{i\sin{\phi(k)\over 2}\cr
\cos{\phi(k)\over 2}}.
\end{equation}
Clearly the free chiral field correspond to
$\phi(k)=\phi_{0}(k)\equiv{\pi\over 2} {k\over{\mid k\mid}}$.
The perturbative vacuum is defined with respect to these operators.
We next write down the Hamiltonian density in terms of
these operators as
\begin{equation}
{\cal T}^{00}(x)=:\left[-i{\bar\psi(x)}\gamma^{1}\partial_{x}\psi(x)
+g{\bar\psi(x)} \psi(x)\sigma +{1\over2}\sigma^{2}\right]:
\equiv {\cal T}^{00}_{F} +{\cal T}^{00}_{int} +{\cal T}^{00}_{\sigma} ,
\end{equation}
where the notations for the three terms are
obvious and : : denotes normal ordering with respect to
perturbative vacuum. In the following we shall treat
the auxiliary field $\sigma$ as a classical field and
retain the quantum nature of $\psi$ to calculate
the effective potential as a function of $\sigma$
at a finite temperature. We shall now calculate
${\cal T}^{00}$ at space time origin. Using equation
(4.3) and (4.5) we then have e.g.
\begin{equation}
{\cal T}^{00}_F=({\cal T}^{00}_F)_c +({\cal T}^{00}_F)_{\tilde c}
+({\cal T}^{00}_F)_{c^{\dag}{\tilde c}}+({\cal T}^{00}_F)_{{\tilde c^{\dag}}
c},
\end{equation}
where, with $\phi_{0}^{'}={\pi k^{'}/2 \mid k^{'}\mid}$,
\begin{mathletters}
\begin{equation}
({\cal T}^{00}_F)_{c}={1\over2\pi} \int c(k)^{\dagger}c(k{^\prime})
k^{\prime}\sin\left({\phi_{0}+\phi_{0}^{\prime}\over{2}}\right)dk dk^{\prime}
,\end{equation}
\begin{equation}
({\cal T}^{00}_F)_{\tilde c}={1\over2\pi}
\int \tilde c(k){\tilde c^{\dagger}}(k^{\prime})
k\sin\left({\phi_{0}+\phi_{0}^{\prime}\over{2}}\right)dk dk^{\prime},
\end{equation}
\begin{equation}
({\cal T}^{00}_F)_{c^{\dagger}\tilde c}={1\over2\pi}
\int c^{\dagger}(k)\tilde c(k^{\prime})
(-ik^{\prime})\cos\left({\phi_{0}-
\phi_{0}^{\prime}\over{2}}\right)dk dk^{\prime}\eqno(8c)
\end{equation}
and
\begin{equation}
({\cal T}^{00}_F)_{\tilde c^{\dagger}\tilde c}={1\over2\pi}
\int \tilde c^{\dagger}(k)\tilde c(k^{\prime})
(-ik^{\prime})\cos\left({\phi_{0}-
\phi_{0}^{\prime}\over{2}}\right)dk dk^{\prime}.
\end{equation}
\end{mathletters}
For ${\cal T}^{00}$ at space-time origin the other operators are
\begin{mathletters}
\begin{equation}
({\bar \psi}\psi)_{c}={1\over2\pi}\int c(k)^{\dagger}c(k^{\prime})
\cos({\phi_{0}+\phi_{0}^\prime\over2})dk dk^{\prime},
\end{equation}
\begin{equation}
({\bar \psi}\psi)_{\tilde c}={1\over2\pi}\int {\tilde c}(k){\tilde c}
(k^{\prime})^{\dagger}
\cos({\phi_{0}+\phi_{0}^\prime\over2})dk dk^{\prime},
\end{equation}
\begin{equation}
({\bar \psi}\psi)_{c^{\dagger}\tilde c}={1\over2\pi}\int c(k)^{\dagger}
\tilde c(k^{\prime})i
\sin({\phi_{0}-\phi_{0}^\prime\over2})dk dk^{\prime}
\end{equation}
and
\begin{equation}
({\bar \psi}\psi)_{\tilde c^{\dagger}\tilde c}={1\over2\pi}\int
\tilde c(k)^{\dagger}\tilde c(k^{\prime})i
\sin({\phi_{0}-\phi_{0}^\prime\over2})dk dk^{\prime}.
\end{equation}
\end{mathletters}
As in equation (2.1), we construct the trial state for the physical vacuum
through
\begin{equation}
|vac'>=exp\left(\int \chi(k)c(k)^\dagger \tilde c(-k)dk\right)|vac>
\end{equation}
where $\chi(k)$ is the trial function.  In Ref. \cite{hm88} we considered
this model through minimisation of the vacuum expectation value of
${\cal T}^{00}$ which is evaluated in a straightforward manner. This
involved a Bogoliubov transformation, where fresh operators $b(k)$ and
${\tilde b}(k)$ were defined
such that $b(k)$ and ${\tilde b}(k)^{\dagger}$ annihilate
the non-perturbative vacuum. The ansatz function is exactly solvable here,
and in fact we obtain that \cite{hm88}
\begin{equation}
\chi(k)=\frac{ik}{g\sigma}\left[\left(1+\frac{g^2\sigma^2}{k^2}\right)
\right].
\end{equation}
With usual renormalisation, this yields the conventional result \cite{grsnv}.
Furthermore, taking even a `wrong' ansatz function with one parameter yields
results very similar \cite{hm88} to the exact one. This is relevant in the
context of the fact that in any problem of physical importance, we have
to take simple trial functions with only a few parameters, and minimise over
these parameters. In priniciple we should take all possible trial
expressions, unless the energy functional is effectively quadratic in the
trial functions as was the case here.

This example is meant to be a simple test case for the application of the
method. We note that in the present example we did not take a unitary
transformation for the definition of the new vacuum.

\section{\bf {Goldstone Theorem}}
We shall now discuss Goldstone theorem \cite{glstn}, which is the most
important result for phase transitions in quantum field theory.
We consider in particular a quantitative link between the restructuring of
vacuum in field theory and the Goldstone particles \cite{nambu} as
localised vacuum modes for bound states dependant on the
vacuum structure \cite{hm92} through explicit constructions.

Let G be a continuous symmetry group of dimension $n$ with generators $Q_i$
$(i=1\cdots n)$. Let H be the Hamiltonian. When symmetry is present,
for $i=1,\cdots,n$,
\begin{equation}
\left[ H,Q_i \right]=0
\end{equation}
and
\begin{equation}
Q_i|vac>=0 .
\end{equation}
If $|vac>$ is unstable, then it will destabilise to a $|vac'>$ as is given
by equation (2.1). Let us have here
\begin{equation}
H|vac'>=E_{vac}|vac'>,
\end{equation}
where, with translational invariance, $E_{vac}=\epsilon _0 \times V$
with $V$ as the
total volume  and $\epsilon_0$ being the energy density for the preferred
vacuum configuration.  When symmetry is broken,
\begin{equation}
Q_i|vac'>\neq 0
\end{equation}
for some $i=1,\cdots,k$ as corresponding to $k$ symmetry breaking directions.
 From equations (5.1) and (5.3) we then obtain that
\begin{equation}
(H-E_{vac}) Q_i|vac'>=0 .
\end{equation}
Thus $Q_i|vac'>$ have zero total energy for $1 \leq i \leq k$, and these are
the Goldstone modes of Goldstone theorem. The extra thing that has been
achieved here is to identify the corresponding states as eigenstates of
the total Hamiltonian. If the vacuum state for vacuum destabilisation is
determined variationally through the minimisation of energy, then the above
equation will yield approximate eigenstates of the Hamiltonian as resulting
from vacuum structure.

A further clarification here shall be useful. We have
\begin{equation}
Q_i=\int J^0_i(\vec x)d \vec x,
\end{equation}
where $J^{\mu}_i(x)$ is the conserved current. Hence we easily identify that
$J^{\mu}_i(\vec x)|vac'>$ creates a Goldstone mode at $\vec x$, and, equation
(5.4) describes the mode with zero total momentum. In fact, this way one
can construct Goldstone modes of finite momentum by replacing equation (5.4)
by
\begin{equation}
|i,\vec p>=\int e^{i\vec p\cdot\vec x}J^0_i(\vec x)d \vec x \;|vac'>.
\end{equation}
Another remark may be pertinent. For a real number $\alpha$, let us consider
the state
\begin{equation}
e^{Q_i\alpha}|vac'>\neq 0 .
\end{equation}
This also has zero eigenvalue corresponding to the $|vac'>$ configuration.
This is however not a Goldstone mode since we can never generate a state
of finite momentum corresponding to it.
This is one of many possible degenerate vacuum configurations not corresponding
to any observable particle, in contrast to that in equation (5.6) which
can be detected. The parallel situation obtains if we have discrete
symmetries, where as above we can not generate particles of finite momenta.
Also, instead of equation (5.7), let us consider the `new' state
\begin{equation}
|i,\vec p>=\int e^{i\vec p\cdot\vec x}J^0_i(\vec x)d \vec x \;U(\alpha)|vac'>,
\end{equation}
where $U(\alpha)$ is some element of G. This again amounts to a redefinition
of $|vac'>$, and will give no new physics. Instead, it will be equivalent
merely to a redefinition of the original operators.

The above features get nicely integrated in the usual proof of Goldstone
theorem through poles of the propagators. We have considered here the
same with states, where the theorem remains the same, but we have the
extra advantage of the bound states (for condensate modes of symmetry
breaking) becoming closely associated with the vacuum structure.

The generators for which $Q_i|vac'>=0$ will form a subalgebra and
define the residual symmetry group S.
These do not give rise to zero modes since $Q_i$ operating on
$|vac'>$ as above do not define states.

We shall now consider examples of the above with specific symmetry
groups as well as the types of symmetry breaking.
Before proceeding to more general cases which bring out the distinctive
features of the present approach,
let us first consider the familiar case with O(3) symmetry.  The Lagrangian
with a triplet of real scalar fields here is given as
\begin{equation}
{\cal L}(x)=\partial _{\mu}\phi_i \partial ^{\mu}\phi_i - {\cal V}(\phi)
\end{equation}
where we have the double well potential with
\begin{equation}
{\cal V}(x)=-\frac{\mu^2}{2}\phi^2 + \frac{\lambda}{4}(\phi^2)^2.
\end{equation}
When we minimise the above potential classically, we obtain the solution
that $\phi^2=\mu^2/\lambda\equiv v^2$ yields a minimum. We can also look at
the same situation as a quantum phase transition with
\begin{equation}
|vac'>=U|vac>,
\end{equation}
where, parallel to Ref \cite{hm88,hgs} we take
\begin{equation}
U=e^{i\int \dot \phi_3(x) d \vec x}.
\end{equation}
In the above, we have taken the symmetry breaking direction as the z-axis.
We also have \cite{hgs}
\begin{equation}
U\phi_3(x)U^\dagger  \equiv \phi_3'(x)=\phi_3(x)-v
\end{equation}
which is equivalent to the substitution $\phi_3=v+\phi_3'$ with $v$ as
the `classical' background field.
Constructing the generators we can then easily see that the charges
$Q_1$ and $Q_2$ with the states given as
$Q_1|vac'>$ and $Q_2|vac'>$ now respectively give the $\phi_2$ and
$\phi_1$ quanta at rest as the zero modes.

In the above only a known mechanism has been quoted. We next consider
the example of chiral symmetry breaking \cite{nambu}
where some new insight is gained.
For this purpose we consider global chiral symmetry breaking in the
context of pions being the Goldstone modes, which will illustrate
the present concepts.

\section{\bf{Chiral Symmetry Breaking}}

Let us consider chiral symmetry for massless quarks with the Lagrangian
given by \cite{nambu}
\begin{equation}
{\cal L}=\bar\psi(i\gamma^{\mu}\partial_{\mu})\psi+G\left[(\bar\psi \psi)^2
+(\bar\psi i \tau _a \gamma^5\psi)^2\right],
\end{equation}
which is invariant under the chiral transformations
\bigskip
\begin{equation}
\psi\rightarrow \psi'=e^{i\gamma^5\tau_a\epsilon_a}\psi.
\end{equation}
In the above we have taken
$\psi=\left(\begin{array}{c}u\\ d \end{array}\right)$ quark doublet
along with a N-plet of $SU(N)$ colour, $G$ is a dimensionful coupling constant,
and the above is a theory with a cutoff $\Lambda$.

We next consider the field expansion
\begin{equation}
\psi(\vec x)=\frac{1}{(2\pi)^{3/2}}\int d \vec k
\left[U_r(\vec k)c_r(\vec k)e^{i\vec k \cdot\vec x}
+V_r(\vec k)d_r(\vec k)^\dagger  e^{-i\vec k \cdot\vec x}\right],
\end{equation}
where, for free fields,
\begin{mathletters}
\begin{equation}
U_r(\vec k)=\frac{1}{\sqrt{2}}\left(\begin{array}{c}1\\ \vec\sigma\cdot\hat k
\end{array}\right)
u_{Ir},
\end{equation}
and
\begin{equation}
V_r(\vec k)=\frac{1}{\sqrt{2}}\left(\begin{array}{c}\vec\sigma\cdot\hat k \\ 1
\end{array}\right)
v_{Ir}.
\end{equation}
\end{mathletters}
The perturbative vacuum is defined through
\begin{equation}
c_r(\vec k)|vac>=0 \; ;\qquad d_r(\vec k)|vac>=0 \; .
\end{equation}
We consider a trial vacuum given as
\begin{equation}
|vac'>=U|vac>\equiv e^{B^\dagger  - B}|vac>,
\end{equation}
where,
\begin{equation}
B^\dagger =\int f(\vec k) u_{Ir}\vec \sigma \cdot \hat k v_{Is}
c_r(\vec k)^\dagger d_s(-\vec k)^\dagger   d \vec k.
\end{equation}
The energy density functional is given as, with $\kappa=|\vec k|$,
\begin{eqnarray}
\epsilon &=&<vac'|{\cal H}(\vec x)|vac'>\nonumber \\
&=&-\frac{2N}{(2\pi)^3}\int^{\Lambda} \kappa cos(2f(\vec k))d\vec k
-\frac{2GN(2N+1)}{(2\pi)^3}\int^{\Lambda} sin(2f(\vec k))d\vec k .
\end{eqnarray}
Through functional differentiation, the above leads to the solution that
\begin{equation}
tan(2f(\vec k))=\frac{2GI(2N+1)}{\kappa},
\end{equation}
where
\begin{equation}
I=\frac{1}{(2\pi)^3}\int^{\Lambda} sin(2f(\vec k))d\vec k \equiv
\frac{M}{\kappa}\;
{\rm say} .
\end{equation}
This leads to the $``$gap" equation
\begin{equation}
M=\frac{2G(2N+1)}{(2\pi)^3}\int^{\Lambda} \frac{M}{\sqrt{M^2+\kappa^2}}d\vec k
{}.
\end{equation}
We can easily verify that the above equation has a solution with
$M \ne 0$ provided
\begin{equation}
G\Lambda^2(2N+1) > 2 \pi^2,
\end{equation}
and that for such a solution, the energy density is lower than the
$``$perturbative" solution with $M=0$.

The new field operators and spinors corresponding to $|vac'>$ shall be denoted
by primes and are related to the old ones through the transformations
\begin{eqnarray}
\psi'(\vec x)&=& U\psi(\vec x)U^\dagger  \nonumber \\
&=& \frac{1}{(2\pi)^{3/2}}\int d \vec k
\left[U_r'(\vec k)c_r'(\vec k)e^{i\vec k \cdot\vec x}
+V_r'(\vec k)d_r'(\vec k)^\dagger  e^{-i\vec k \cdot\vec x}\right],
\end{eqnarray}
where, now we have the new spinors given as
\begin{mathletters}
\begin{equation}
U_r'(\vec k)=\frac{1}{\sqrt{2}}\left(\begin{array}{c}cos(f(\vec k))
+sin(f(\vec k))\\
(cos(f(\vec k))-sin(f(\vec k)))(\vec\sigma\cdot\hat k) \end{array}\right)
u_{Ir},
\end{equation}
and
\begin{equation}
V_r'(\vec k)=\frac{1}{\sqrt{2}}\left(\begin{array}{c}(cos(f(\vec k))-sin(f(\vec
k)))
(\vec \sigma \cdot \hat k)\\
cos(f(\vec k))+sin(f(\vec k))\end{array}\right)
v_{Ir},
\end{equation}
\end{mathletters}
which may be recognised as a Bogoliubov transformation.
We may now easily see that as already stated the pions with zero total
momentum as Goldstone modes are now given as
\begin{equation}
|\pi_a(\vec 0)>=N_{\pi}\int sin(2f(\vec k))q'(\vec k)^\dagger
\tau  _a \tilde q'(-\vec k) d \vec k
|vac'>,
\end{equation}
where $N_{\pi}$ is a normalisation constant, and, $q(\vec k)^\dagger $ and
$\tilde q(-\vec k)$ are the creation operators respectively for the
isodoublet quark
and antiquark pairs. We note that the extra result of the present statement
of the Goldstone theorem with explicit vacuum realignment leading to
phase transition yields the pion wave function, which quantitatively depends on
the vacuum structure. We are thus able to see the pion as a local zero mode
disturbance of the vacuum.

Further details regarding this have been considered elsewhere \cite{hm92}.

\subsection{\bf{Goldstone Theorem and Approximate Symmetry Breaking}}
It is clear that if the symmetry was initially not exact, then Goldstone
theorem is not applicable to predict the existence of zero modes.
However, what if the symmetry were almost exact? To examine this situation
with the above picture, let us have
\begin{equation}
H=H_s+H_{sb}
\end{equation}
where $H_s$ maintains the symmetry and $H_{sb}$ is the symmetry breaking
part assumed to be small. For the case of continuous symmetry,
as before let there be a phase transition with $Q_i$ being a generator
corresponding to a symmetry breaking direction. In that case we can
take the first order perturbation theory corresponding to the above
Hamiltonian, and suggest that the mass of the Goldstone mode is given
as, with $H=H_{eff}+E_{vac}$,
\begin{equation}
m_i\big|_{\rm Goldstone}=\frac{<vac'|Q_i H_{eff}Q_i]|vac'>}
{<vac'|Q_i Q_i|vac'>}.
\end{equation}
We note that in the above $Q_i|vac'>$ creates a state of zero total
momentum, and hence both the numerator and denominator above shall contain
$\delta (\vec 0)$, which shall cancel.

As a particular case let us consider chiral symmetry breaking for the pion
as below. We first note that for the numerator we may write,
\begin{eqnarray}
&& Q_aH_{eff}Q_a\equiv Q_a[H_{eff},Q_a] \equiv -[H_{eff},Q_a]Q_a\nonumber\\
&\equiv & \frac{1}{2}[[Q_a,H_{eff}],Q_a]=\frac{1}{2}[[Q_a,H_{sb}],Q_a].
\end{eqnarray}
Let us consider the case of chiral symmetry breaking with explicit
symmetry breaking term as
\begin{equation}
H_{sb}=m\int \psi(\vec x)^\dagger \gamma^0\psi(\vec x)d\vec x
\end{equation}
with $m$ as the current quark mass. We can then easily evaluate that
we now have with
\begin{eqnarray}
Q_a&=&\int J^5_a(\vec x)d\vec x \nonumber \\
&=&\int \psi(\vec x)^\dagger \gamma ^5 t_a\psi(\vec x) d \vec x,
\end{eqnarray}
\begin{equation}
<vac'|Q_aH_{eff}Q_a|vac'>\equiv -\frac{1}{2}m<vac'|\bar\psi\psi|vac'>V,
\end{equation}
where V is the total volume associated with $\delta (\vec 0)$ mentioned
earlier.

We now simplify the denominator in equation (6.17). Here
the pion decay constant is given through
\begin{equation}
<vac'|J^5_a(x)|\pi_b(\vec p)>=\delta _{ab}\frac{f_{\pi}}{(2\pi)^{3/2}}
\frac{p^0}{\sqrt{2p^0}}.
\end{equation}
Saturating the denominator with pion states, this yields that
\begin{eqnarray}
<vac'|Q_aQ_a|vac'>&=&\int <vac'|Q_a|\pi_b(\vec p)>d\vec p<\pi_b(\vec p)
|Q_a|vac'> \nonumber\\
&=&\frac{m_{\pi}}{2}f_{\pi}^2 V.
\end{eqnarray}
In the above, as earlier, there is no summation over the index $a$, but
the index $b$ is summed. Equations (6.17), (6.21) and (6.23) now
yield the well-known formula
\begin{equation}
m_{\pi}^2=-\frac{m<\bar \psi \psi>}{f_{\pi}^2}
\end{equation}
of current algebra, which is really what has been carried out above,
in the context of explicit vacuum destabilisation. We note that in the
presence of the current quark, $Q_a|vac'>$ is not the exact pion state,
and with the present picture, the correction due to this can be calculated.
The above formula is correct to the extent that the pion wave function
can be completely described through vacuum destabilisation, and that
the denominator in equation (6.17) can be saturated with pion states only.
Also, in contrast to the conventional use of Goldstone theorem
through the pole of the propagator, in the above we are able to
explicitly construct the state, and hence can obtain the wave function
for the same as well as predict its size, which gets
related to the scale dimensions for the destabilised vacuum.

\section{\bf Phase transition in Quantum Chromodynamics}
\bibcite{schwinger}{14}
\bibcite{schut85}{15}
In the present section we shall consider phase transition in quantum
chromodynamics with gluon condensates. For this purpose, we
start with the QCD Lagrangian given as
\begin{equation}
{\cal L} =-{1\over 2}G^{a\mu\nu}(\partial_{\mu}{W^{a}}_{\nu}
-\partial_{\nu}{W^{a}}_{\mu}+gf^{abc}{W^{b}}_{\mu}{W^{c}_{\nu})}
+{1\over 4}{G^{a}}_{\mu\nu}{G^{a\mu\nu}},
\end{equation}
\noindent where ${W^{a}}_{\mu}$ are the SU(3) color gauge fields.
We shall quantise in Coulomb gauge and write the
electric field  ${G^{a}}_{0i}$ in terms of
the transverse and longitudinal parts as
\begin{equation}
{G^a}_{0i}=
{^TG^a}_{0i}+{\partial_{i}{f}^{a}},
\end{equation}
\noindent where the form of ${f}^{a}$ is to be determined.
In the Coulomb gauge the
subsidiary condition and the equal time algebra for the gauge
fields are given as \cite{schwinger}
\begin{equation}
{\partial_{i}}{W^a}_i=0
\end{equation}
\noindent and
\begin{equation}
\left [{W^{a}}_{i}({\vec x},t),^{T}{G^{b}}_{0j}
({\vec y},t)\right ]=i{\delta}^{ab}({\delta}_{ij}
-{{\partial_{i}\partial_{j}}\over {\partial^2}})\delta
({\vec x}-{\vec y}).
\end{equation}
\noindent We take the field
expansions for ${W^{a}}_{i}$ and $^{T}{G^{a}}_{0i}$
at time t=0 as \cite{schut85}
\begin{mathletters}
\begin{equation}
{W^{a}}_{i}(\vec x)={(2\pi)^{-3/2}}\int
{d\vec k\over \sqrt{2\omega(\vec k)}}({a^a}_{i}(\vec k) +
{{a^a}_{i}(-\vec k)}^{\dagger})\exp({i\vec k.\vec x})
\end{equation}
 \noindent and
 \begin{equation}
 {^{T}G^{a}}_{0i}(\vec x)={(2\pi)^{-3/2}} i \int
{d\vec k}{\sqrt{\omega(\vec k)\over 2}}(-{a^a}_{i}(\vec k) +
{{a^a}_{i}(-\vec k)}^{\dagger})\exp({i\vec k.\vec x}).
\end{equation}
\end{mathletters}
\noindent From equation (7.4) these give the commutation
relations for ${a^a}_{i}$ and ${{a^b}_{j}}^\dagger$ as
\begin{equation}
\left[ {a^a}_{i}(\vec k),{{a^b}_{j}(\vec k^{'})}^\dagger\right]=
\delta^{ab}\Delta_{ij}(\vec k)\delta({\vec k}-{\vec k^{'}}),
\end{equation}
\noindent where, $\omega$(k) is arbitrary \cite{schut85}, and,
 \begin{equation}
 \Delta_{ij}(\vec k)={\delta_{ij}}-{k_{i}k_{j}\over k^2}.
\end{equation}
\noindent In Coulomb gauge, the expression for the Hamiltonian
density, ${\cal T}^{00}$ from equation (7.1) is given as \cite{schwinger}
\begin{equation}
{\cal T}^{00}=:{1\over 2}{^{T}{G^a}_{0i}}{^{T}{G^a}_{0i}}+
{1\over 2}{W^a}_{i}(-\vec \bigtriangledown^2){W^a}_{i}+
{{g^2}\over 4}f^{abc}f^{aef}{W^b}_{i}{W^c}_{j}{W^e}_{i}{W^f}_{j}+
{1\over 2}(\partial_{i}f^{a})(\partial_{i}f^{a}):,
\end{equation}
\noindent where : : denotes the normal ordering with respect to
the perturbative vacuum, say $\mid vac>$, defined through
${a^a}_{i}(\vec k)\mid vac>=0$. In equation (7.8) we have not retained
the cubic terms in gauge fields {\it since with gluon pair condensates
these will not contribute}. The term ${1 \over 2}(\partial _if^a)
(\partial _if^a)$ automatically includes interactions for both
time-like and longitudinal gluons, here through the auxiliary field
description, and ${\cal T}^{00}$ is calculated after this elimination
through equations (7.9) and (7.10) given later.
We note that \cite{am91}
\begin{equation}
f^a=-{W^a}_0-g \; f^{abc}\;{ (\vec \bigtriangledown ^2)}^{-1}
({W^b}_i \; \partial _i {W^c}_0).
\end{equation}
\noindent Hence, eliminating $f^a$, the equation for ${W^a}_0$ becomes
\begin{equation}
\vec \bigtriangledown ^2{W^a}_0 + 2g \; f^{abc}{W^b}_i \; \partial _i
{W^c}_0 + g^2 \; f^{abc}f^{cde} \; {W^b}_i \partial _i
((\vec \bigtriangledown ^2)^{-1}({W^d}_j \partial _j{W^e}_0))
={J^a}_0,
\end{equation}
\noindent where
\begin{equation}
{J^a}_{0}=gf^{abc}{W^b}_{i}^{T}{G^c}_{0i}.
\end{equation}
\bibcite {tfd}{16}
It is not possible to solve the above equation
for ${W^a}_0$.
We shall therefore proceed with a mean field type of approximation
 for the ground state. In thermofield method \cite{tfd}
 the ground state is written as $\mid vac';\beta>$ with $\beta=1/T$ and
 it is a state in an extended Hilbert space including thermal modes. We
 shall therefore replace the left hand side of equation (7.10)
 with the expectation values for $\mid vac';\beta>$ for all
the fields other than ${W^a}_0$. Then, the above equation gets replaced by
\begin{eqnarray}
\vec \bigtriangledown ^2{W^a}_0 (\vec x )
&&+ g^2 \; f^{abc}f^{cde} \;<vac',\beta\mid  {W^b}_i(\vec x ) \partial _i
(\vec \bigtriangledown ^2)^{-1}({W^d}_j(\vec x ) \mid vac',\beta>
\partial _j{W^e}_0(\vec x ))\nonumber\\ && ={J^a}_0(\vec x ).
\end{eqnarray}

\bibcite{hmtlk}{17}
At zero temperature, $\mid vac';\beta=\infty>=\mid vac'>$
was the nonpertubative ground state as discussed in \cite{am91}.
We shall consider here only this aspect of the problem for quantum
chromodynamics \cite{am91}. The modifications at finite temperature
as an application of thermofield dynamics will be discussed separately
in \cite{hmtlk} in this symposium.
We define  $\mid vac'>$ through a unitary
transformation, in a similar manner to
Gross-Neveu model considered earlier \cite{hm88}, given as
\begin{equation}
\mid vac ^{'}>
=U\mid vac>,
\end{equation}
 \noindent where
 \begin{equation}
 U=\exp({B^\dagger}-B),
\end{equation}
The unitary operator U for the temperature dependent case
will be discussed later.
It was shown  that \cite{am91} at zero temperature, we may have
\begin{equation}
{B^\dagger}={1\over 2}
\int {f(\vec k){{{a^a}_{i}(\vec k)}^\dagger}
{{{a^a}_{i}(-\vec k)}^{\dagger}}d\vec k},
\end{equation}
\noindent
where $f(\vec k)$ describes gluon condensates. It was also
shown \cite{am91} that a nonperturbative vacuum with $f(\vec k)\not =0$
was favoured above a critical coupling.
Temperature dependence will arise with $U$ above depending
on temperature and the Hilbert space being doubled.
For a moment we continue to take zero temperature.
With the above transformation, the operators, say
${b^a}_{i}(\vec k)$, which annihilate $\mid vac^{'}>$ are
given as
\begin{mathletters}
\begin{equation}
{b^a}_{i}(\vec k)=U{a^a}_{i}(\vec k){U}^{-1}.
\end{equation}
\noindent We explicitly evaluate from the above equation
the operators ${b^a}_{i}(\vec k)$
corresponding to the state, $\mid vac^{'}>$ as related to the
operators corresponding to the state, $\mid vac>$ through
the Bogoliubov transformation given as
\begin{equation}
\left(
\begin{array}{c}
 {b^a}_{i}(\vec k) \\ {{b^a}_{i}(-\vec k)}^\dagger
 \end{array}
 \right)
=\left(
\begin{array}{cc}
 \cosh f(\vec k) & -\sinh f(\vec k) \\
-\sinh f(\vec k) & \cosh f(\vec k)
\end{array}
\right)
\left(
\begin{array}{c}
{a^a}_{i}(\vec k)  \\
{a^a}_{i}(-\vec k)^\dagger,
\end{array}
\right)
\end{equation}
\end{mathletters}
\noindent where the function
 $f(\vec k)$  is even in $\vec k$ and has been assumed to be real.
\noindent Using equations (7.6) and (7.16b), we obtain the same
commutation relation for the operators ${b^a}_{i}$
and ${{b^b}_{j}}^\dagger$ given as
\begin{eqnarray}
\left[ {b^a}_{i}(\vec k),
{{b^b}_{j}(\vec k^{'})}^\dagger \right]
& = & \delta^{ab}\Delta_{ij}(\vec k)\delta({\vec k}-{\vec k^{'}}),
\nonumber
\end{eqnarray}
 which merely reflects that the Bogoliubov
transformation (16b) is a canonical transformation.
It is useful to define
\begin{equation}
<vac' \mid :{W^a}_{i}(\vec x){W^b}_{j}(\vec y):\mid  vac'>=
{\delta }^{ab}
\times (2  \pi )^{-3}\int d\vec k e^{i\vec k.(\vec x-
\vec y)}\; {F_{+}(\vec  k)\over \omega (k)}\;
\Delta _{ij}(\vec k),\end{equation}
\begin{equation}{<vac' \mid}: {^{T} G^{a}_{0i}} (\vec x),
{^{T} G^{b}_{0j}} (\vec y):{\mid vac'>} = \delta ^{ab} (2 \pi )^{-3}
\int d{\vec k}e^{i{\vec k}.{(\vec x-\vec y)}}
{\Delta _{ij}(\vec k)\over\omega (k)} F_{-}( k).\end{equation}
In the above $F_{+}(k)$ and $F_{-}(k)$ are given as
\begin{mathletters}
\begin{equation}F_{+}(\vec k)=\biggl (
{\sinh 2f(k)\over 2}
+ {\sinh}^{2}f(k)
\biggr ),\end{equation}
\begin{equation}F_{-}(\vec k)=\biggl ({\sinh}^2 f(k)
- {\sinh 2f(k)\over 2}
\biggr ).\end{equation}
\end{mathletters}
Using the above it was earlier seen that the solution for
$\tilde W^a \! _0 (\vec k)$ is given by
\begin{equation}
(\vec k ^2+ \phi (\vec k ))\tilde W^a \! _0(\vec k )=
-\tilde J^a \! _0(\vec k ),
\end{equation}
 \noindent where
$\tilde W^a \! _0 (\vec k)$ and $\tilde J^a \! _0 (\vec k)$
are Fourier transforms of $W^a \! _0 (\vec x)$
and $J^a \! _0 (\vec x)$; and, with $\;{ g^2/ 4 \pi }=\alpha _s$,
\begin{equation}\phi (\vec k )=
{3 \alpha _s \over 2 \pi } \int {F_+(k')\over {\omega (k')}}dk'
\Big[ (k^2+k'^2)-
{ (k^2-k'^2)^2
\over 2 kk'}\times ln \Big |{k+k'
\over k-k'} \Big | \Big].
\end{equation}
 \noindent We note that $f(\vec k)$ occuring in the Bogoliubov
 transformations describing the possible vacuum structure is to
 be determined through minimisation of energy density at zero
 temperature. We may achieve this in simple cases \cite{hm88}, but here it
 is impossible to do so. Hence we adopt the alternative procedure of
 taking a reasonably simple ansatz for $f(\vec k)$ and extremise it
 over parameters in $f(\vec k)$. Since the gluon correlation function
 $f(\vec k)$ should go to zero for large k (condensation being a long
 distance effect) and since in the above equation hyperbolic
functions enter, we choose the simple form, with $k=\mid\vec k\mid$,
\bibcite {svz}{18}
 \begin{mathletters}
 \begin{equation}
 \sinh f(\vec k)=Ae^{-Bk^{2}/2},\end{equation}
\noindent where the parameter $A$ is determined through
energy minimisation and the dimensional parameter $B$
gets determined by the SVZ parameter \cite{svz}.
We note that we then have from equation (7.16b)
\begin{equation}
<vac'|{a^a}_i(\vec k)^\dagger {a^a}_i(\vec k')|vac'>=8\times 2\times A^2
e^{-B\kappa ^2} \delta (\vec k-\vec k'),
\end{equation}
\end{mathletters}
so that we are really taking a Gaussian distribution for the {\it perturbative}
gluons in physical vacuum as a natural ansatz. Here we are aiming at
an understanding of the
vacuum structure in the context of low energy phenomenology, including the
properies of the same at finite temperature in one framework. It is clear that
such a solution will be applicable only for problems involving low energy
physics, as is our objective for the study of phase transitions.

 Using the form (21) for $f(\vec k)$, the energy density,
$\epsilon_{0}$ can be written in terms of the dimensionless
quantities $x={\sqrt {B}}k$ and $\mu={\sqrt B}m$ as \cite{am91}
\begin{eqnarray}
\epsilon_{0}
& = & {1\over {B^2}}(I_{1}+I_{2}+{I_{3}}^{2}+I_{4})\nonumber \\
& & \equiv {1\over {B^2}}F(A),
\end{eqnarray}
\noindent where
\begin{mathletters}
\begin{equation}
I_{1}
={4\over {\pi^2}}\int {\omega(x)x^{2}dx
(A^{2}e^{-x^2}-Ae^{-{{x^2}/2}}{(1+A^{2}e^{-x^2})}^{1\over 2})},
\end{equation}
\begin{equation}
I_{2}=
{4\over {\pi^2}}\int {{x^{4}}\over \omega(x)}dx
(A^{2}e^{-x^2}+Ae^{-{{x^2}/2}}{(1+A^{2}e^{-x^2})}^{1\over 2}),
\end{equation}
\begin{equation}
I_{3}=
{{2g}\over {\pi^2}}\int {x^{2}\over \omega(x)}dx
(A^{2}e^{-x^2}+Ae^{-{{x^2}/2}}{(1+A^{2}e^{-x^2})}
^{1\over 2}),\end{equation}
\noindent and
\begin{equation}
I_{4} =4\times {(2 \pi )^{-6}}\int d{\vec x}
{G(\vec x)\over {x^2+\phi (x)}}
\end{equation}
\end{mathletters}
\noindent with
\begin{eqnarray}
 G(\vec x) & = & {3g^{2}}
\int d \vec x'
\biggl (A^{2}e^{-{x'}^2}+Ae^{-{{x'}^2}/2}
{(1+A^{2}e^{-{x'}^2})}^{1\over 2}\biggr )\nonumber \\ & \times
& \biggl (A^{2}e^{-{(\vec x+\vec x')}^2}-
Ae^{-{{(\vec x +\vec x')}^2}/2}
{(1+A^{2}e^{-{(\vec  x+\vec x')}^2})}^{1\over 2}\biggr )\nonumber \\
& \times &
{\omega (\mid \vec x +\vec x'\mid)\over {\omega ( x')}}
\biggl (1 +{{{({x'}^2 +\vec x .\vec x')}^2} \over
{\vec {x'}^2 (\vec x + \vec x')^2}}\biggr ) ,
\end{eqnarray}
\noindent  and
\begin{eqnarray}
 \phi  (x) & = & {3g^2\over  {8 \pi  ^2}}\int dx'
 \biggl (  x^2+{x'}^2-{(x^2-{x'}^2)^2
\over{2xx'}}\log \Big | {{x+x'}\over{x-x'}}\Big | \biggr )
\nonumber \\ & \times & \biggl (A^{2}e^{-{x'}^2}+Ae^{-{{{x'}^2}/2}}
{(1+A^{2}e^{-{x'}^2})}^{1\over 2}\biggr ).
\end{eqnarray}
\noindent In the above, $\omega(x)={(x^2+\mu^2)}^{1\over 2}$,
with $\mu$ as the effective gluon mass in ${1\over {\sqrt B}}$
units .
We then identify the gluon mass $\mu$ from
the sum of the single contractions of the
quartic interaction term of ${\cal T}^{00}$ in equation (7.8),
the negative of which gives a mass term in the effective Lagrangian.
We thus have the $self consistency$ $ requirement$ that
\begin{equation}
\mu^2={{2g^2}\over {\pi^2}}
\int {{x^2}dx\over \omega(x)}
(A^{2}e^{-x^2}+Ae^{-{{x^2}/2}}{(1+A^{2}e^{-x^2})}
^{1\over 2}).
\end{equation}
\noindent $\mu$ is determined through an iterative procedure
for any particular value of A so that equation (7.27)
 is satisfied with the input $\mu$ on the right hand side
 giving rise to the same output $\mu$ on the left hand side.
We found that there exists
a critical value $g_c$ of $g$ with ${{g_c}^{2}}/4\pi\simeq 0.39$,
such that for $g > g_{c}$, $A_{min}\not= 0$ and
$\epsilon_{0}$ becomes negative demonstrating instability of
perturbative vacuum.
Two remarks here may be in order. Firstly we note that we have
taken here $g$ as the coupling constant in the Lagrangian.
In semiperturbative QCD, through renormalisation group equation
the coupling constant of the Lagrangian together with the scale
for renormalisation generate the mass scale $\Lambda$ of QCD,
which describes the running coupling constant. At the present
level of only discussing the vacuum energy such a mathematical
structure has not been arrived at. For this purpose, we have to consider
higher order Greens functions with nonperturbative vacuum structure
so that we are in a position to consider the renormalised coupling
constant at different scales and see how it runs. The problem is
nontrivial and has not been addressed here. The second remark we
wish to make is the fact that $g_c > 0$ may be an artifact of the
approximation scheme. We encountered such a situation for Gross-Neveu
model in Ref \cite{hm88}.

 The value of B, which is a scale parameter,
is now determined by relating it with the SVZ parameter.
In fact, the vacuum structure
 of QCD is given as \cite{svz}
\begin{equation}
{{g^2}\over {4{\pi}^2}}<:{G^a}_{\mu\nu}
{G^{a\mu\nu}:>_{vac^{'}}=0.012GeV^{4}},
\end{equation}
\noindent where $\mid vac^{'}>$ is the physical
vacuum. The left hand side of the above equation can be
explicitly evaluated as
\begin{equation}
{{g^2}\over {4{\pi}^2}}<:{G^a}_{\mu\nu}
{G^{a\mu\nu}:>_{vac^{'}}=
{1\over {B^2}}{{g^2}\over {\pi^2}}
{(-I_{1}(A)+I_{2}(A)+I_{3}(A)^{2}-I_{4}(A))\Big |
_{A=A_{min}}}},
\end{equation}
\noindent where $A_{min}$ is the value of A
corresponding to minimised energy density
and using equation (7.28), the scale parameter, $B$
gets determined as a function of coupling constant.

We can also study the effect of temperature on the
above expressions.  For this purpose, the methodology
of thermofield dynamics \cite{tfd},
with the $``$thermal vacuum" $\mid vac^{'};\beta>$
being defined so as to yield correct distributions for bosons and
fermions shall be useful \cite{hmtlk}. This consists in the doubling of the
Hilbert space, introducing fresh $``$tilde" space with operators
${{\tilde b}^a}_{i}(\vec k)$ and
${{{\tilde b}^a}_{i}(\vec k)}^{\dagger}$ for this space \cite{tfd}
and is discussed in \cite{hmtlk}.

We note that one feature of the present method is obtaining the mass-like
parameter for the gluon fields (in Coulomb gauge) in a
nonperturbative manner which gets defined  through
equation (7.27) with self-consistency requirement \cite{am91}.
The mass parameter $\mu$ occurs both as an input as well as an output
in that equation, reminding us of Schwinger Dyson integral equations for
self-energy.

\section{\bf Vacuum Destabilisation and Particle Production}
We shall show here how the present picture of phase transition can
lead to particle production with vacuum destabilisation. For this
purpose we consider Salam Weinberg symmetry breaking and then examine
the possibility of Higgs particle production with local vacuum
destabilisation.  We shall also consider some cosmic ray events
like $``$chirons", $``$halos" and muon anomalies from directions of
Cygnus X3, Hercules X1 in terms of local vacuum destabilisation \cite{hgs}.

\subsection {\bf Salam Weinberg Symmetry Breaking}
Let us first consider
here the quantum description of the
conventional spontaneous symmetry breaking (SSB)
for U(1) symmetry. The Lagrangian here is
\begin{equation}
{{\cal L}=(D_{\mu}{\phi})^{\star}(D^{\mu}{\phi})
-V({\phi})-{1\over 4}F_{{\mu}{\nu}}F^{{\mu}{\nu}}}
\end{equation}
with
\begin{equation}
{V({\phi})=-m^{2}{\phi}^{\star}{\phi}+{\lambda}({\phi
}^{\star}{\phi})^{2}.}
\end{equation}
${D_{\mu}={\partial}_{\mu}-igA_{\mu}}$
is the covariant derivative. Classically, when $m^{2}>0$
symmetry breaking occurs with $<{\phi}^{\star}{\phi}>=
{m^{2}\over{2{\lambda}}}$. To give a quantum description to
SSB of Higgs mechanism, let us
write ${\phi}={1\over{\sqrt 2}}{({\phi}_{1}+i{\phi}_{2})}$
with ${\phi}_{1}$ and ${\phi}_{2}$ as real fields. Thus
the equal time algebra
\begin{equation}
{\left[{\phi}({\vec {x}},t),{\dot{\phi}}({\vec {y}},t)^{*}
\right]=i{\delta({\vec {x}}-{\vec{y}}})}
\end{equation}
is equivalent to
\begin{equation}
{\left[{\phi}_{i}({\vec {x}},t),{\dot{\phi}}_{j}({\vec {y}},t)
\right]=i\delta_{ij}{\delta({\vec {x}}-{\vec{y}}})}.
\end{equation}
This equal time algebra is  for interacting fields
as it holds good for interacting operators \cite{hgs,schut85} and
is consistent with the expansions
\begin{mathletters}
\begin{equation}
\phi_{1}({\vec {x}},0)={1\over{\sqrt{2\lambda_{x}}}}{(a(
{\vec {x}})+a({\vec {x}})^{\dagger})}
\end{equation}
and
\begin{equation}
\dot\phi_{1}({\vec {x}},0)=i\sqrt{\lambda_{x}\over2}(-a(
{\vec {x}})+a({\vec {x}})^{\dagger})
\end{equation}
\end{mathletters}
with a and $a^{\dagger}$ satisfying the commutation
relation
\begin{equation}
{\left[a({\vec {x}}),a({\vec {y}})^{\dagger}\right]=
\delta(\vec {x}-\vec {y}),}
\end{equation}
and, for free felds,
$\lambda_{x}=(-\bigtriangledown_{x}^{2} +m^{2})^{1\over 2}$,
but here it could be arbitrary. The Lagrangian (1) has been normal ordered
with respect to $\mid vac>$ defined by $a(\vec{x})\mid vac>=0.$
Let us now define a coherent state $\mid vac^{\prime}>$ as \cite{spm87,hm88}
\begin{eqnarray}
\mid vac^{\prime}>&=&\exp \left(\xi\int{\sqrt{\lambda_
{z}\over 2}}\left(a(\vec {z})^{\dagger}-a(\vec {z})\right)
d\vec {z}\right)\mid vac>\nonumber \\
&\equiv & U \mid vac>,
\end{eqnarray}
where U is a unitary operator. We wish to define
annihilation operator corresponding to $\mid vac'>$, such that
\begin{equation}
\phi_{1}^{\prime^{an}}(\vec {z})\mid vac^{\prime}>=0 .
\end{equation}
Clearly we then have
\begin{eqnarray}
\phi_{1}^{\prime^{an}}(\vec {z})&\equiv&{ 1\over{
\sqrt {2\lambda_{z}}}}a^{\prime}(\vec {z})=
 U \phi_{1}^{an}(\vec {z})U^{-1}\nonumber \\
&=&\phi_{1}^{an}(\vec {z})-{\xi\over 2}.
\end{eqnarray}
Hence for the complex field $\phi$, the vacuum expectation value
(VEV) is given as
\begin{equation}
<vac^{\prime}\mid\phi(\vec {z})\mid vac^\prime>={\xi
\over\sqrt 2}.
\end{equation}
The expectation value of the Hamiltonian density,
using equations (7.1), (7.2) and (7.7), is given as
\begin{equation}
\epsilon_{0}=<vac^{\prime}\mid{\cal T}^{00}\mid vac^
\prime>$$
$$=-{1\over{2}}m^{2}\xi^{2}+{\lambda\over{4}}\xi^{4}
.\end{equation}
Minimisation of energy density $\epsilon_0$
with respect to $\xi$ now gives the usual result
\begin{equation}
\xi\equiv\xi_{0}=\left(m^2\over{\lambda}\right)^{1/2},
\end{equation}
which makes $\epsilon_{0}$ negative for
$m^2>0$, thus demonstrating that $\mid vac^\prime>$
is the ground state. This gives a description of
phase transition as a vacuum realignment at  quantum level, which we
shall use later to discuss Higgs production
with $\underline {local}$ vacuum excitations. Let us now reorder
the Lagrangian (1) with respect to $\phi^{\prime}$
operators. Then we have e.g.,
\begin{equation}
:{(D_{\mu}\phi)}^{\star}{(D^{\mu}\phi)}:=N_{\phi^\prime}
\left((D_{\mu}\phi^\prime)^{\star}(D^{\mu}\phi^\prime)
\right)+{{g^2\xi^2}\over{2}}A_{\mu}A^{\mu}$$$$+{ig\xi\over
{\sqrt{2}}}\left[A_{\mu}(\partial^{\mu}\phi^{\prime})
-A_{\mu}(\partial^{\mu}\phi^{\prime^\star})\right],
\end{equation}
where $N_{\phi^\prime}$ denotes normal
ordering with respect to $\phi^\prime$
operators. Thus gauge bosons now have a mass $g\xi_{0}$.
Similarly, mass of the physical scalar field becomes as usual
$\sqrt{2\lambda}\xi$.
 Thus conventional Higgs mechanism is reproduced
through reordering with respect to the new quantum
state $\mid vac^\prime>$.

We now consider Salam-Weinberg theory. Here
\begin{eqnarray}
{\cal L}&=&-{1\over 4}F_{\mu\nu}^{a}F^{a\mu\nu}
-{1\over 4}B_{\mu\nu}B^{\mu\nu}+{\mid D_{\mu}
\phi\mid}^{2}+{{\overline\psi}_{L}{\gamma}^{\mu}
{\cal D_{\mu}}\psi_{L}}+{\overline e_{R}}\gamma^{\mu}
(i\partial_{\mu}-g^{\prime}B_{\mu})e_{R}\nonumber \\
&-&{\lambda_{1}({\overline{\psi}_{L}}\phi e_{R}
+{\overline e_{R}}\phi^{\dagger}\psi_{L})
-V(\phi).}
\end{eqnarray}
In the above,
\begin{equation}
V(\phi)=-m^{2}\phi^{\dagger}\phi+\lambda{(\phi^{\dagger}
\phi)}^{2},
\end{equation}
with $\phi$ being a complex scalar doublet
given as
\begin{equation}
\phi=\pmatrix{\phi^{(+)}\cr\phi^{(0)}}.\end{equation}
For the present description of the Higgs mechanism,
we may now identify $\phi^{(0)}$ of equation (8.16)
with $\phi$ of equation (8.1)
and run through the algebra as quoted.
We then obtain that $\mid vac^{\prime}>$ defined in
equation (8.7) is the description of electroweak
theory with SSB and $<vac^{\prime}\mid\phi\mid vac
^{\prime}>=\pmatrix{0\cr {\xi\over \sqrt 2}}$.
The tree level potential is then given by, parallel to (8.11)
\begin{equation}
V(\xi)=-{1\over{2}}m^{2}\xi^{2}+{\lambda\over{4}}\xi^{4}
.\end{equation}
We can then see that the present gauge covariant
kinetic term
 $\mid D_{\mu}\phi\mid^{2}$ with
respect to $\mid vac^{\prime}>$ as in equation (8.13),
now yields the usual masses of the gauge bosons and Higgs
particles as
\begin{equation}
m_{W^\pm}={g\xi_{0}\over 2},\quad {m_{Z^{0}}=
{{(g^{2}+{g^{'}}^{2})}^{1/2}\over 2}\xi_{0},\quad
m_{H}=\sqrt {2\lambda}\xi_{0},}\end{equation}
where, $\xi^{0}={\sqrt {m^{2}\over{\lambda}}}$ is the field expectation
value corresponding to minimum of the potential as in equation (8.17).
We note that the above results are identical with the results for
Higgs mechanism starting with an equation parallel to equation (8.9).

\subsection {\bf Temperature Dependence of Vacuum}
We shall now consider the  temperature dependance of $\mid vac'>$,
where, the above non-perturbative
solution of field theory corresponds to the zero
temperature. This is clear when we note that
for any operator $\hat O$,
\begin{equation}
\lim_{\beta\rightarrow\infty}{{Tr(\exp {(-\beta H)}\hat O)}
\over{Tr(\exp{(-\beta H)})}}={<vac^{\prime}\mid\hat O\mid vac
^{\prime}>\over{<vac^{\prime}\mid vac^{\prime}>}},\end{equation}
\noindent where $\mid vac^{\prime}>$ now is the {\it state
of lowest energy} corresponding to zero temperature. We shall
now generalise this to finite temperatures. For this purpose
we shall use the methodology of thermofield dynamics \cite{tfd,hmtlk}.
The idea is to calculate effective potential
at finite temperature as an expectation value over a
$``$thermal vacuum" which at zero temperature reduces
to equation (8.11). Thermal vacuum is given by
\begin{equation}
\mid vac^{\prime},\beta>=U(\beta)\mid vac^{\prime}
>\end{equation}
with
\begin{equation}
U(\beta)=\exp{(B^{\dagger}-B)},\end{equation}
where, thermal modes are created with \cite{tfd}
\begin{equation}
B^{\dagger}=\int\theta(k,\beta){a(\vec k)^{\prime}
}^{\dagger}\tilde {a}{(-\vec k)}^{\dagger}
d\vec k\end{equation}
In the above the operator
$\tilde {a}$ $(\tilde {a}^{\dagger})$ are the
annihilation (creation) operators in the extra
Hilbert space and satisfy the same quantum algebra.
We note that we consider the thermal excitations over
$\mid vac^{'}>$, the stable vacuum after Salam Weinberg
phase transition at T=0.
 For bosons, the function $\theta (k,\beta)$ is
given by \cite{tfd}
\begin{equation}
\sinh^{2}\theta(k,\beta)={1\over {\exp ({\beta
\omega (k,\beta))}-1}}.\end{equation}
corresponding to the Bose distribution
of the number operator
$a'(\vec k)^{\dagger}a'(\vec k)$ for the
physical particles and commutator algebra.
The Bogoliubov transformation corresponding
to the thermal vacuum becomes
\begin{equation}
\pmatrix{a(\vec k,\beta)^{\prime}\cr
\tilde a(-\vec k,\beta)^{\dagger}}=\pmatrix{
\cosh{\theta(k,\beta)} & -\sinh{\theta(
k,\beta)}\cr-\sinh{\theta(k,\beta)}
& \cosh{\theta(k,\beta)}}{\pmatrix{a(\vec k)^{\prime}\cr
\tilde a(-\vec k)^{\dagger}}}.\end{equation}

A few remarks regarding construction of
thermal vacuum as in equation (8.20) may be relevant.
Clearly, at zero temperature,
with $\beta=\infty$, the Hilbert space consisting of
only physical particle decouples in the extended
Hilbert space. Further,
$\mid vac^{\prime},\beta>$ contains Higgs particle
excitations as is clearly seen from equations (8.21) and (8.22).
In principle we should include the gauge bosons
and fermion excitation as well in $B^{\dagger}$. As
a first approximation we  ignore these
channels
to illustrate the qualitative features and possible
experimental signatures with only the Higgs sector.

As is clear from the Lagrangian in equation (8.14),
and the fact that $\mid vac^{\prime},\beta>$
contains only $\phi_{1}{^{\prime}}$ quanta,
the contribution to energy density
will come from $\phi_{1}^{\prime}$
dependent terms only when expectation value of the
Hamiltonian with respect to thermal vacuum is
taken.We thus have the expression for the
{\it effective Hamiltonian} density ${\cal T}_{eff}^{00}$ as
\begin{equation}
{\cal T}_{eff}^{00} ={1\over{2}}\left[{\dot \phi_{1}^{\prime
^{2}}}+\phi_{1}^\prime(-\vec\bigtriangledown^{2})\phi_{1}^\prime+
{\lambda\over{2}}\phi_{1}^{\prime^{4}}+
(3\lambda{\xi^{2}}-m^{2})\phi_{1}^{\prime^{2}}+
{\lambda\over {2}}\xi^{4}-m^{2}\xi^{2}\right].\end{equation}
The energy density at temperature $\beta$ now becomes
\begin{eqnarray}
V(\xi,\beta)&\equiv &\epsilon(\beta)={<vac^{\prime},\beta\mid
{\cal T}^{00}_{eff}\mid vac^{\prime},\beta>}\nonumber \\
=&&{1\over 2}.(2\pi)^{-3}\int{(\omega(k,\beta)^{2}+\vec k^{2}
+3\lambda{\xi^2}-m^2)\over {\omega(k,\beta)(\exp(\beta\omega
(k,\beta))-1)}}d\vec k\nonumber \\
+&&{{3\lambda}\over 4}\left[(2\pi)^{-3}\int{1\over{\omega(k,\beta)
({\exp(\beta\omega(k,\beta)}-1)}}d\vec k\right]^{2}
+{\lambda\over 4}\xi^4-{m^2\over 2}\xi^2,\end{eqnarray}
where
\begin{equation}
\omega(k,\beta)=\sqrt{\vec k^2+m_{H}(\beta)^{2}},\end{equation}
with $m_{H}(\beta)$ being
the Higgs mass at temperature $1/\beta$.
Equation (26) is an implicit definition
for the effective potential with the parameters to be determined
self-consistently as explained below.

In thermofield method, temperature dependent field theory
needs the physical mass of the Higgs particle as input
while using the distribution function in the calculation of
temperature dependent effective potential as in
the right hand side of equation (8.26) through $\omega(k,\beta)$.
At finite temperatures Lorentz invariance is broken
and thus mass is not well defined.
We shall however still follow the method of defining mass
through second derivative of effective potential at its minimum
along with a self consistency requirmentnt.
The mass of the physical Higgs particles shall
be taken as
${d^{2}V(\xi,\beta)/ d\xi^{2}}\mid_{\xi=\xi_{min}}$
where $\xi_{min}$ corresponds to the expectation value of the
field for the minimum of effective potential
 as on the left hand side of equation (8.26).
 However for the calculation of $V(\xi,\beta)$ in equation (8.26)
 we need a value for the mass like parameter $m_{H}(\beta)$
 on right hand side. We shall call  the
mass on the right hand side of equation (8.26) for such
calculations as the input
mass and the square root of the expression ${d^{2}V(\xi,\beta
)/ d\xi^{2}}\mid_{\xi=\xi_{min}}$ after $V(\xi,\beta)$ is evaluated
as the output mass.
Self-consistency demands that both be the same.
Such a determination of mass through self-consistency is very much like
solving the integral equation for self energy in perturbative
calculations with self-energy both as an input as well as output.
 The situation here is
however technically different in the sense that
effective mass enters on the right hand side
of equation (8.26) for using thermal distributions, and on the left
hand side through second order derivative of potential at its minimum.
We determine $m_{H}(\beta)$ in equation (8.26) through an iterative
procedure until input mass equals the output mass.
The methodology here is not a loop calculation, but self-consistency
simulates
the dynamical effects of something like loop calculations.
We note that
this definition, however, is not the same as through the pole
of the propagator and, alternative definitions of mass might give
somewhat different results.

For numerical evaluations it is
useful to rewrite equation (8.26) in terms of the
dimensionless quantities with the substitutions
\begin{equation}
z={\xi\over\xi_{0}} , \quad \mu={m_{H}(\beta)\over \xi_{0}}\quad
{\rm and}\quad y=\beta\xi_{0},\end{equation}
where $\xi_0=\sqrt{m^2/\lambda}$ is the value
of $\xi_{min}$ for {\it zero temperature}.  The expression for effective
potential now becomes
\begin{eqnarray}
V(z,y)&=&\xi_{0}^{4}\left[{\lambda\over 4}z^4-{\lambda\over 2}z^2
+{1\over 2}I_1(z,y)+{3\lambda\over 4}(I_2(z,y))^2\right]\nonumber \\
&&{\equiv{\xi_{0}^{4} V_{1}(z,y)}},\end{eqnarray}
where
\begin{mathletters}
\begin{equation}
I_{1}(z,y)={1\over 2{\pi}^2}{\int\limits_0^{\infty}{{x^2
(\omega(x)^2+x^2+\lambda(3z^{2}-1))}\over {\omega(x)(\exp{(y\omega(x))}
-1)}}}dx
\end{equation}
and
\begin{equation}
I_{2}(z,y)={1\over 2{\pi}^2}
\int\limits_{0}^{\infty}{{x^2}\over{\omega(x)(\exp{(y\omega(x))}-1)}}
dx
\end{equation}
with
\begin{equation}
\omega(x)=\sqrt{x^{2}+\mu^{2}}.\end{equation}
\end{mathletters}
In the above $\mu$ is the Higgs particle mass in $\xi_{0}$
units and as stated is to be determined self-consistently.
V(z,y) as a function of $\xi$ and $\beta$
is accepted as the effective potential at finite temperature
 only after an iterative determination of
$\mu$ or $m_{H}(\beta)$ through self-consistency.
The results of the calculations
are stated below.

\bibcite{doljack}{19}
In fig.1 we have plotted $V(\xi,\beta)-V(0,\beta)$ for
different temperatures. For $T=T_{c}\approx 2.1\xi_{0}$, the
shape of the potential changes, which thus determines the
critical temperature $T_{c}$, which is the same as obtained by
Dolan and Jackiw \cite{doljack}.  In fig.2 we have then plotted in curve I
$m_{H}(\beta)^{2}$ as a function of temperature $T$.
For $T=T_{c}$,  $m_{H}(\beta)^{2}$ goes to zero as expected. It however
again rises for temperatures $T > T_{c}$.
We have also plotted in curve II of the same figure
$m^{2}(\beta)\equiv {d^{2}V(\xi,\beta)/ d\xi^{2}}
\mid_{\xi=0}$
corresponding to the $``$mass-square" of the effective Lagrangian at
$\xi=0$ for the double well
potential, which as expected is negative below $T_{c}$ and becomes the same
as $m_{H}(\beta)^{2}$ above $T_{c}$ as $\xi_{min}=0$.
We have next plotted  the order parameter $\xi_{min}$ as a
function of temperature in curve III of the same figure. The order
parameter goes to zero as temperature approaches the critical
temperature $T_{c}$.  $T_{c}$ could
be determined through any of these three curves
 and turns out to be ${2.1\xi_0\simeq 525GeV}$.
We did not use here  loop expansion or a high
temperature approximation \cite{doljack} but used the thermofield method with
a self consistent variational ansatz giving results similar to those
of Ref \cite{doljack}.
It is reassuring that the present nonperturbative numerical technique
yields the same result inspite of the expressions being different.
 The gap in energy density of the thermal vacuum
with respect to the  vacuum at zero temperature is given by
\begin{eqnarray}
\Delta\epsilon(\beta)&=&V(\xi_{min},\beta)-V(\xi_{min},\beta=\infty)\nonumber
\\
&=&\xi_{0}^4\left[V_{1}(z_{min},y)+{\lambda\over 4}\right].\end{eqnarray}
The number density of Higgs particles
at temperature ${\beta}$ is here given as
\begin{eqnarray}
N(\beta)&=&<vac^{\prime},\beta\mid a(z)^{\dagger}
a(z)\mid vac^{\prime},\beta>\nonumber \\
&=&{(2\pi)^{-3}}{\int {1\over {{\exp{(\beta\omega(k,\beta))}-1}}}
d\vec {k}}.
\end{eqnarray}

\subsection {\bf Local Destabilisation of Vacuum}
\bibcite{gunton}{20}
We shall now consider possible local $``$heating" of the vacuum with
particle collision,which, from the very nature of the
problem, will be intuitive and heuristic and can only lead to
qualitative conclusions. As in
condensed matter physics, such a destabilisation of
vacuum can occur if we pump in enough energy into
a small macroscopic volume which thermalises $\it locally$
and becomes hot. The dynamics of such a $``$bubble" formation is
known to be  complex \cite{gunton} and shall not be tackled here.
We shall rather consider the possibility of excitation of
vacuum with a bubble formation which has a nonzero temperature
and the thermal equilibrium is local. With vacuum as the medium in
which collisions take place, we conjecture here that during  particle
collisions a part of the collision energy might excite vacuum locally.

We shall now apply the above ideas to such a situation with
temperature defined locally inside the bubble. Thus the energy density
gap is no longer constant throughout the volume but is maximum at
center of collison and decreases
to zero away from it. The total energy of such a locally excited
region or the bubble shall be given as
\begin{equation}
E_{B}=\int\Delta\epsilon(\beta(
\vec r))d\vec r,
\end{equation}
where $\Delta\epsilon(\beta (\vec r))$ is the same function as in equation
(8.32) except that $\beta$
is spatially dependant. Similarly the number of Higgs particles
inside the bubble becomes
\begin{equation}
N_{B}=\int{N(\beta({\vec{r}}))d\vec {r}}.
\end{equation}
As stated the determination of $\beta(\vec r)$
as a function of $\vec r$ from first principles
is impossible. We shall therefore look into the qualitative
structure only through a gaussian distribution for the same.
For the temperature distribution inside the
bubble we shall therefore take
\begin{equation}
\beta(r)^{-1}=T(r) =T_{0}\exp{(-ar^2)}.
\end{equation}
In the above, $T_{0}$ is the temperature
at the centre of bubble and the parameter 'a' essentially decides the
region over which vacuum is excited with the corresponding $``$volume"
being $\approx a^{-{3\over 2}}$.

\bibcite{keil}{21}
We have noted in equation (8.32) that such a bubble will contain
Higgs particles. Such particles in Salam Weinberg theory
are however coupled to fermions and get converted to quark or lepton pairs
\cite{keil}
as the bubble cools. The Higgs particles in the bubble will primarily go
to heavy fermion anti-fermion pairs.  The decay width of such
a process for free Higgs particles at finite temperature is given as
\cite{keil}
\begin{equation}
\Gamma(\beta)=\tanh\left({{\beta m_{H}(\beta)}\over 4}\right)
\Gamma_{0},\end{equation}
where $\Gamma_{0}$ is the decay width for $H\rightarrow f\bar f$
at zero temperature. Then the average decay width of the
Higgs particles in the bubble may be calculated to be
\begin{equation}
\Gamma_{avg}=\Gamma_{0}\int \tanh\left({{\beta m_{H}(\beta(r))}\over
4}\right)N_{B}(\beta (r))d\vec r. \end{equation}
Inside the bubble, the masses of the fermions decrease
in the same manner as the VEV or mass of the Higgs particles, since
the masses of the fermions are generated through the vacuum
expectation values. As they come out of the bubble, these will
become more massive as the temperature outside decreases resulting
in cooling of the bubble.  The average mass of the
{\it decaying} Higgs particles will depend on both their life times
as well as the number of Higgs particles with a specific
mass. Thus we may calculate this average mass as
\begin{eqnarray}
M_{avg}&=&\int N_{B}(\beta (r)){\Gamma(\beta(r))\over \Gamma_{avg}}m_{H}
(\beta(r))d\vec r\nonumber \\
&=&{{\int N_{B}(r)\tanh({{\beta m_{H}(\beta(r))}\over 4})m_{H}(\beta(r))
d\vec r}\over{\int N_{B}(\beta (r))\tanh({{\beta m_{H}(\beta(r))}\over 4})
d\vec r}}\end{eqnarray}
{\it  which will correspond to the average transverse momentum
of the decaying particles} at any fixed time. This however will be a
dynamic phenomenon as the bubble cools here mainly through the decay
of Higgs particles with corresponding loss of energy. To get an idea of the
mass
distribution inside the bubble we may calculate the variance of the same as
\begin{equation}
\sigma^{2}=\left[(M^{2})_{avg}-(M_{avg})^{2}\right],\end{equation}
where $(M^{2})_{avg}$ is given as
\begin{equation}
(M^{2})_{avg}=\int N_{B}(\beta (r)){\Gamma(\beta(r))\over \Gamma_{avg}}m_{H}
(\beta(r))^{2}d\vec r.\end{equation}
We note that the conjecture quoted in equation (8.35)
for the temperature distribution in a bubble has not been so far utilised.
We shall now explicitly take it to get a qualitative idea of possible
experimental signals. For this purpose, the total energy as in
equation (8.33) should be avilable for vacuum excitation from some
collision process. Further, the number of Higgs particles in the
$``$bubble" as in equation (8.34) should be large enough
 for thermal equilibrum to be possible.
With this in mind, we now look for qualitative experimental signals.

\subsection {\bf Experimental Signatures}
The experimental signatures of bubble formation as compared
to $``$heating" of any material in condensed matter physics
is extremely sharp, since the bubble here has energy dissipation
through the conversion of Higgs particles to fermion and antifermion
pairs, which has no earlier parallel. The signature
here is particularly strong since there will be a {\it preferential}
production of heavy fermion and antifermion pairs, as the coupling of
fermions to Higgs particles is proportional to their masses. Such a
{\it preferential} production of heavy fermions is  absent
for the conventional mechanisms for particle production. Hence,
irrespective of the details of the modelling, excessive heavy fermion
production in a collison
can be a signature for bubble formation. The explicit modelling
will decide whether we could anticipate to observe the same
anywhere, or interpret existing data with the association
of Higgs particles.
Before doing this, however, we shall note some preliminary calculations
for different $E_B$ and $``$a" in equations (8.33) and (8.35) as inputs.
We shall consider the possibility of a destabilisation
of vacuum over a volume $\approx a^{-{3\over 2}}$ due to a
collision. A natural ansatz here appears to be the length scale
associated with the cross-section of the collision process.
Hence, for vacuum excitation, we shall tentatively assume
that $\sigma \simeq {\pi \over a}$, and compute the possibility
of signals. It is quite possible that such a hypothesis be not
correct. If so, the scales quoted below will change, and the process
may be visible in a different way or not visible at all.
This assumption is made for conceptual convenience and
seem to correspond to some signals in cosmic rays.

For definiteness let us consider a cross-section of
the order of microbarns,which corresponds
to a value $0.02\xi_{0}^2$ for 'a' in equation(8.35). The corresponding
bubble energy versus number of Higgs particles is
plotted for $\lambda \simeq 0.02$ corresponding to Higgs mass of 50 GeV
in fig.3 as curve I. The results are not very sensitive
to $\lambda$ except for kinematics. We observe
that the multiplicity increases almost
 linearly
with bubble energy for $E_{B}$ larger than one TeV.
We note that for $E_{B}$ greater than $2 \sim 3$ TeV, there are
 reasonable number of Higgs particles.The
possibility of vacuum excitation will depend upon the fraction
of the total energy of collision that goes as $E_{B}$ for local
excitation and bubble formation. What is this fraction and how it
depends on energy is not known. In curve II of fig.3 we plot the
central temperature against bubble energy $E_{B}$ for $N_{B}$
around and larger than thirty since thermal effects for small $N_{B}$
will not be sensible. We note that for temperatures of order
150 GeV, which is much less than the critical
temperature of 525 GeV, the number of Higgs
particles appears to be sizable as may be seen in fig.3.
We note here that with inclusion of gauge particles and
fermions $T_c$ could become smaller than $525$ GeV. In that case, the
present mechanism could show up at a lower temperature. This is because
$N_B$ will become appreciable at a lower temperature as the Higgs mass
at that temperature will go down compared to the present case.

The bubble energy will depend on the fraction of collision energy that
goes towards vacuum excitation. This is not known. However, we note
from fig.3 that for $E_{B}$ around 2 TeV or more, the above process
could become relevant.
Curve I here gives the multiplicity for on mass shell
Higgs particles, which is temperature dependent.
Such a bubble will primarily lose energy
and cool rapidly through Higgs paricles getting converted to fermion pairs
dynamically, thus using up bubble energy.
This production of fermions, shall include the fact that
the fermion masses are proportional to the Higgs
masses above. As the fermions come out,
the bubble cools and the fermions
become more massive.
The characteristic feature of such a process leading to Higgs
particle decay yields e.g. more high $P_{T}$ muon pairs
and an overall excess of muons due to heavy flavor production.
This will be larger than the expected probability
for them because of preferential heavy flavor production for Higgs
particle $``$decay".
Such events shall be rare, but, could be just visible in
accelerator experiments beyond the TeV range.
As stated these will have
unusually high multiplicitiy and an excess of muons, which shall
signify the onset of the above mechanism.
We note that the above comments are qualitative, but shall leave
visible signatures inspite of the dynamics not being
understood.

Since as per the present calculations, the accelerator energies are
inadequate, we state below some signatures of this type in cosmic rays.
For this purpose let us examine the following events which
 appear not to be capable of being explained with simulation programmes
of known physics.

\noindent {\bf (i)  Chiron events:}

\bibcite{chiron}{22}
These events \cite{chiron} were
observed in Brasil-Japan collaboration in Chacaltaya Emulsion Chamber
experiment in chambers 19 and 21 and have the following characteristics.
The total $P_{T}$ is of the order of 5 GeV there. Further inside
them we have $``$mini-clusters" associated with as small a $P_{T}$
as $\simeq$ 10 MeV-20 MeV. These, we note, may be associated
with Higgs particle production as above. In the present scenario,
as may be seen from equation (8.38), the high $P_{T}$
of the order of 5 GeV will correspond to decay of
the Higgs particle. These particles will produce heavy flavors
and hence the $``$mini clusters" of 10-20 MeV transverse momentum
might indicate $D^{*}$ production. Production of $``$mini clusters"
along with absence of $\pi^{0}$ could thus be interpreted
as signature of Higgs particle productions in these events.

One more observation in this context may be relevant. The total
transverse momentum of the chirons will correspond to the mass
of the Higgs particle \cal {when it decays}. Higgs particles
could decay during the process of cooling of the bubble even before
zero temperature is reached. Thus the present identification
of chiron events where large $P_{T}$ comes from the decay
of Higgs particles to heavy fermion pairs does not determine its mass.
\newpage
\noindent {\bf (ii) Halo events:}

\bibcite{halo}{23}
Halo events \cite{halo} in cosmic rays are some events with
excessive multiplicity
where the particle number could not be counted \cite{halo}. As is clear from
the
$E_{B}$ versus $N_{B}$ plots, the multiplicity of Higgs particles
rises almost linearly with the bubble energy. Hence, for energies beyond the
threshold for the present process, the multiplicity \cal { will
increase linearly with energy which is much
faster than what can be expected from ordinary physics}. Hence
the $``$halo" events \cite{halo} may indicate the onset of new physics through
an unusual rise in the multiplicity resulting from excitation of
vacuum. Further,the multiple cores in halos \cite{halo} just look like multiple
bubble formation in condensed matter physics \cite{gunton}!

\noindent {\bf (iii)Cygnus X3, Hercules X1 signals:}

\bibcite{weekes}{24}
\bibcite{halzen}{25}
These are signals
from extremely high energy air showers originating from the
directions of Cygnus X3 and of Hercules X1
which have a high muon content and thus indicates {\it hadronic}
interactions. Similar results from Crab nebula have also been suggested
\cite{weekes}.It is then presumed that a neutral {\it stable
hadronic particles} e.g. Cygnets, quark nuggets etc. from Cygnus
X3 and Hercules X1 should be coming. Instead, we propose that
{\it local vacuum excitations}
with Higgs particle production as above could give rise to
the excess muon signals through preferential heavy flavour
production \cite{halzen}.

The present mechanism of Higgs particle production consists of
three parts. (a) The bubble will get formed; (b) the temperature
dependent vacuum will have a quantum mechanical description with
production of off-mass-shell Higgs particles and (c) the dissipation
of the bubble takes place through particle production via conversion of
Higgs particles to fermion pairs. In the present
note we have concentrated on (b) i.e. the description of
temperature dependent local excitation of vacuum. It is possible to
obtain a signal for the process because of {\it preferential} heavy
flavor production during the dissipation of
the bubble. The processes (a) and (c) are not only non-perturbative
but also time dependent and an understanding of the same
is a non-trivial step \cite{gunton}.  However, heavy flavor production
for (c) leaves a trail which can be always seen if we can look
for this.

A comment regarding thermalisation may be relevant.
Production of a $``$coherent bubble" as in equation (8.7) may be
instantaneous corresponding to a new value $\xi(\beta)$. However,
time scales for local thermalisation of vacuum is not known and
will be associated with bubble formation \cite{gunton}.
Once the existence
of the process is confirmed with more observations, the time scales
involved for this process for thermalisation, the response of vacuum
regarding thermal and transport properties for the same as a medium
needs to be investigated and understood.

\bibcite{ferrer}{26}
\bibcite{goldberg}{27}
We note that with conventional consideration of field theory
with high temperature approximation a careful and complete calculation
has been done by Ferrer et al \cite{ferrer} where three possible phases
have been identified, including one with W-condensates.
Our calculations are complementary in the sense that high
temperature limit is not taken and that experimental
signatures at temperatures less than critical temperature
could be observed. We may further observe that a recent analysis
by Goldberg with scalars only shows the instability of the perturbative
tree level calculations regarding cross-sections at ultra high energies
\cite{goldberg}.
In this context or otherwise, the nonperturbative techniques
developed here may be more relevant and we may look for signatures of
heavy flavor production or unusual rise in multiplicity.

 Thus, to sum up, the characteristic features of Higgs particle
production through vacuum destabilisation would be (i) relatively high
$P_{T}$ for Higgs particle decay, (ii) preferential production of heavy
flavors and (iii) an unusual rise in multiplicity with energy as a
result of this mechanism. We find that the observations
of Chirons, $``$halo" as well as the hadronic signal associated with
the direction of Cygnus X3 and Hercules X1 could indicate
the onset of the present mechanism, and observation of the same
in accelerators may be possible in not too far a future. It is significant to
note that even though Salam Weinberg phase transiton takes place
around a temperature of more than 500 GeV,
the effects of local destabilisation of vacuum could be seen
for temperatures as low as 150 to 200 GeV , with the corresponding
bubble energies being a few TeV.

We describe the above in the context of phase transitions in quantum
field theory because this clearly illustrates the conceptual differences
as well illustrates the extra experimental consequences.
We may also emphasize the nonperturbative manner for the definition of
the Higgs particle mass through a self-consistency requirement in
equation (8.26), similar to the solution of equation (7.26) of the
last chapter.

\section{\bf{Discussions}}
Normally we think that quantum mechanics supersedes classical mechanics,
and quantum field theory supersedes quantum mechanics. Is quantum phase
transition a finer description of phase transition compared to the same
in classical mechanics and including the same? To understand this,
it may be worthwhile to consider phase transitions in condensed matter
physics.

Let us consider a crystalline solid where quantum corrections are
relevant. The base state
here may be recognised as the nuclei at specific mean lattice sites.
The quantum modes above it may be imagined as consisting of the thermal
vibrations of the nuclei, as well as the states for the valence and/or
conduction electrons. With a change of temperature, the mean positions
of the lattice sites may slightly alter. This will generally happen in
a continuous manner, and corresponds to a destabilisation of the base
state, but not a phase transition. However, at a critical temperature, this
can happen in a discontinuous manner. For example, a cubic lattice
may change its configuration. This as per our definition will correspond
to a quantum phase transition. The expectation values of some operators
will have a discontinuity reflecting the change in the base state.
This can be treated classically in some
cases, but dealing with it in quantum mechanical (or field theoretic)
fashion can also be done, and can have more predictive power.
The present techniques could be useful for some of these problems
\cite{anderson}.

With a simpler modelling, the change in the base state above with
temperature may be imagined as the temperature dependance of the fermi
surface of a solid. This will normally happen in a continuous manner;
however, at a critical temperature, this change will be discontinuous.
This as earlier will give rise to a discrete change in the expectation
values of some quantum operators indicating a phase transition.

\bibcite{nmtr}{28}
\bibcite{dtrn}{29}
 Parallel concepts have also been used for nuclear matter at zero
 as well as finite temperatures \cite{nmtr} with a dressing of the
nuclear medium by scalar isoscalar pion pairs. We may even picture bound
states \cite{spm87} where the medium or `vacuum' is destabilised locally.
This has been used for example to consider deuteron \cite{dtrn} as
an almost realistic problem, again with pion dressing.

We thus note
that the present nonperturbative approach not only is physically
appealing, but also deals with a wide class of problems in a unified
manner. It becomes particularly relevant for
phase transitions where vacuum is constructed to have an explicit structure
which can be tested experimentally.

\newpage
\centerline{\bf Figure Captions}
\vspace{.5in}
\noindent{\bf Figure 1:} Effective potential $V(\xi ,\beta)- V(0,\beta)$ has
been plotted at different temperatures $ T={1\over\beta}$. Critical temperature
 $T_{c}\approx 2.1\xi_{0}\approx 525 GeV$.
 \bigskip

\noindent{\bf Figure 2:} In curves I, II and III we have plotted
$m_{H}(\beta)^2$,
${d^{2}V\over d\xi^2}\mid_{\xi =0}$ and $\xi_{min}(\beta)$ respectively
as functions of temperature in units of $\xi_{0}\approx 250 GeV$.
Critical temperature
$T_{c}$ can be determined from any of the above curves as $2.1\xi_{0}$.
\bigskip

\noindent {\bf Figure 3:} In curve I we plot Higgs particle multiplicity
 $N_{B}$ against bubble
energy $E_{B}$ in units of TeV corresponding to
microbarn size. In curve II
we also plot the central temperature $T_{0}$ upto 250 GeV against bubble
energy $E_{B}$.
\vfil\break
\end{document}